\documentclass{iau}

\usepackage{amsmath}
\usepackage{graphicx}
\usepackage{multirow}

\newcommand{\apj }{ApJ}
\newcommand{\apjl }{ApJL}
\newcommand{\apjs }{ApJS}
\newcommand{\mnras}{MNRAS}
\newcommand{\pasj}{PASJ}
\newcommand{\nat}{Nature}
\newcommand{\aap}{A\&A}
\newcommand{\Msol}{M{$_{\odot}$}}
\newcommand{\per}{$^{-1}$}
\newcommand{\um}{$\mu$m}
\newcommand{\kms}{km~s{$^{-1}$}}
\newcommand{\Vlsr}{$\rm V_{LSR}$}
\newcommand{\Htwo}{$\rm H_{2}$}

\begin{document}

\lefttitle{John Bally}
\righttitle{The CMZ Asymmetries:  Feeding or Feedback?}

\jnlPage{1}{7}
\jnlDoiYr{2021}
\doival{10.1017/xxxxx}

\aopheadtitle{Proceedings IAUS Symposium 405}
\editors{M. Zaja\v{c}ek,  T. Je\v{r}\'{a}bkov\'{a}, V. Karas, R. Schödel \&  P. Sukov\'{a}, eds.}

\title{The CMZ Asymmetries:  Feeding or Feedback? }

\author{John Bally}
\affiliation{University of Colorado, Boulder}

\begin{abstract}
Three-fourths of the dense gas and dust in the CMZ is located at positive longitudes 
and positive radial velocities.  The majority of compact 24 $\mu$m sources are 
at negative longitudes.  These two asymmetries  indicate either a recent asymmetric  
injection of gas along the bar dust lanes,  or that most of the molecular gas is contained 
in a small number of massive, gravitationally bound clouds, or a major feedback episode 
which dissociated an entire sector of the CMZ's dense gas.
\end{abstract}


\maketitle

\section{Introduction}

The central $\sim$300~pc radius region of our Galaxy is dominated by molecular gas \citep{2022MNRAS.516.3911S}.
Three-quarters of this gas is located at positive longitudes and positive radial velocities with respect to the Galactic center; only one-quarter is at negative longitudes \citep{1987ApJS...65...13B,1988ApJ...324..223B,1998ApJS..118..455O,2020MNRAS.498.5936E}.    The gas asymmetry is evident in the spatial distribution of dust traced by diffuse continuum emission between 70 \um\  and 3 mm.    The Milky Way's CMZ is more asymmetric than the central regions of most barred galaxies such as NGC1097 or NGC1300.

Clusters of 24~\um\ sources in the major star-forming complexes such as Sgr B1, B2, and the 50 \kms\ cloud at positive longitudes trace young stellar objects (YSOs).   However, over two-thirds of the point-like 24~\um\ sources observed with Spitzer are located at negative longitudes between $\sc l$ = -1$^o$ and 0$^o$ in a small scale-height population away from molecular clouds \citep{2009ApJ...702..178Y}.   Spectroscopy shows that some are main sequence OB stars surrounded by 24 \um -emitting dust while others are post-main sequence red-giants and supergiants \citep{2011ApJ...736..133A,2015ApJ...799...53K,2018A&A...609A.109N}.           Figure 1 shows the two asymmetries.  

Between $\sim$1 to 300 pc from the supermassive black hole (SMBH) the spherically averaged mass density scales as $\rm \sim R^{-1.8}$.   The equivalent circular orbit speeds vary from $\sim$120 to $\sim$200~\kms .    For an orbit speed of $\rm V_{200}$ = 200~\kms, the orbit time  is $\rm t = 3 R_{100} V^{-1}_{200}$ Myr where $\rm R_{100}$ is the Galactocentric radius in units of 100 pc.      

\section{Short orbit times, shear, and the asymmetries}

I consider three models for the origins of the asymmetries:   
{\bf 1:} Recent asymmetric infall of a large fraction of the CMZ's gas mass.  
{\bf 2:} Stochastic asymmetry resulting from the CMZ's molecular gas being confined to a small number of bound clouds.
{\bf 3:} Disruption of an entire sector of the CMZ by starburst-powered feedback.

{\bf 1: Recent infall?}  \citet{2018MNRAS.475.2383S} and \citet{2019MNRAS.484.1213S} proposed that  infall along the leading edge dust lanes of the Galactic bar is episodic and estimated the  infall rate to be  $\sim 2.7^{+1.5}_{-1.7}$ ~ \Msol yr\per .    
\citet{2020MNRAS.499.4455T} used more advanced simulations to infer an inflow rate of $\sim$1~\Msol  yr\per .  
\citet{2024A&A...689A.121V} found a ~200~pc long, $\rm \sim 2.4 \times 10^6$~\Msol\ stream falling into the CMZ from the near-side bar dust lane implying an infall rate of $\sim 1-2$~\Msol  yr\per\ over the last Myr.  \citet{2024ApJ...971L...6S,2025ApJ...984..109S} traced the mass inflow from the 3 kpc arms into the CMZ, finding an average inflow rate of $\rm \sim$1.1~\Msol yr\per\ over the last 15-20 Myr.   Cloud collisions caused by overshooting from the far dust lanes produce the high velocity-dispersion complexes at $\sc l$= 3.2$^o$  5.0$^o$, and 7.0$^o$.   The mass of gas in the CMZ has been estimated to be  2 - 5$\rm \times 10^7$~\Msol .  At these rates, $\sim$10  to 50  Myr would be required to accumulate the CMZ's mass.   Episodic accretion from mostly one side of the Galactic bar could produce the observed asymmetry in the gas if the infall occurred more recently than several CMZ orbit times.   However, this model does not explain the 24~\um\ asymmetry.   

\begin{figure}
    \includegraphics[scale=0.246]{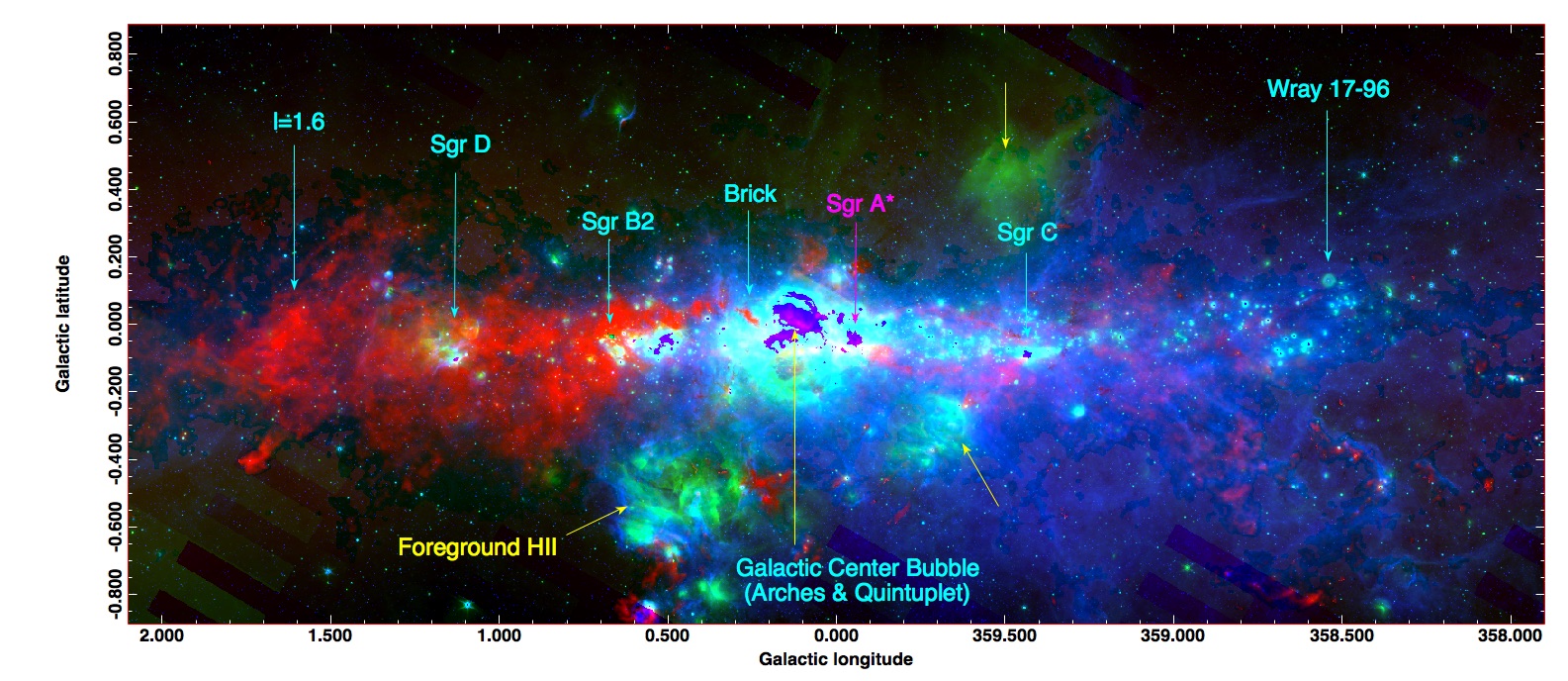}
    \caption{A wide field view of the CMZ showing the Herschel HiGAL column density of dust associated with the dense CMZ molecular gas (red), 24~$\mu$m (green), and 8~$\mu$m (blue) emission traced by Spitzer. }
    \label{sample-figure}
\end{figure}

{\bf 2: Stochastic asymmetry?}  If the CMZ is a long-lived structure present for many orbit times and contains of many dozens of  clouds or gravitationally unbound gas,  shear would re-distribute the gas into a more symmetric configuration in a few orbit times.     Symmetrization can be avoided {\it if} most of the gas is contained in a small number of self-gravitating clouds sufficiently compact and dense to resist tidal disruption.  Such clouds have to be denser than $\rho > 3 V^2(R) / 2 \pi G R^2$ $\approx 3 \times 10^{-20} V^2_{200} R^{-2}_{100}$ (g cm$^{-3}$).   With only a few clouds, the gas asymmetry may be stochastic.    More clouds could be located on one side of the center than the other as seen from the Sun.  But, this model also fails to address the 24~\um\ asymmetry.

{\bf 3: Recent starburst?}   \cite{1984Natur.310..568S} found a degree-scale ($\sim$200 pc) bubble of radio emission expanding orthogonal to the CMZ, the `Galactic Center Lobes' (GCL).    The northeastern wall crosses the plane at the brightest non-thermal filaments (NTFs) near the 3 to 5 Myr old Quintuplet cluster at $\sc l \approx$+0.2$^o$.   Feedback from OB stars in the Quintuplet and younger Arches clusters drives the $\sim$30 pc diameter `Galactic Center Bubble'  (Fig.1) centered at $\sc l \approx$0.15$^o$,  the brightest and most compact super-bubble in the Galaxy.   MeerKAT radio emission, X-rays, and mid-IR warm dust in the GCL trace a chimney over a 420 pc extent above and below the plane \citep{2019Natur.573..235H,2021A&A...646A..66P}.  Its axis of symmetry crosses the Galactic plane at negative longitudes (Fig. 2).  The southwestern wall of the GCL crosses the plane near Sgr C at $\sc l \approx$-0.8$^o$.  Although,  \citep{2026A&A...710A.205K,2019PASJ...71...80N} argue that the GCL is a foreground structure, I suspect this bipolar flow is in the CMZ.

The axis of the GCL  is centered on the cluster of 24~\um\  sources at negative longitudes (Fig. 2).   Molecular gas here is mostly confined to two negative \Vlsr\  branches above and below the  24~\um\ sources  between Sgr A and Sgr C.  Figure 3 shows a closeup of this region.    Many small HII regions, ring-shaped features, some partial shells, and extensive filaments oriented roughly along the Galactic plane are present at 1.28 GHz.    ACES 100 GHz and  H-recombination lines indicate that they trace ionization fronts or sheets and filaments confined by the CMZ's strong magnetic fields \citep{2026ApJ..1000..206C}.   Filaments roughly at right angles to the plane trace NTFs.  A 9 pc by 14 pc ring of molecular emission is centered at [$\sc l,b$] =[359.73$^o$,0.00$^o$] between \Vlsr = -24 to -44~\kms\ (Fig 3, right).  These features trace impacts of stellar feedback.    

\begin{figure} 
    \includegraphics[scale=0.223]{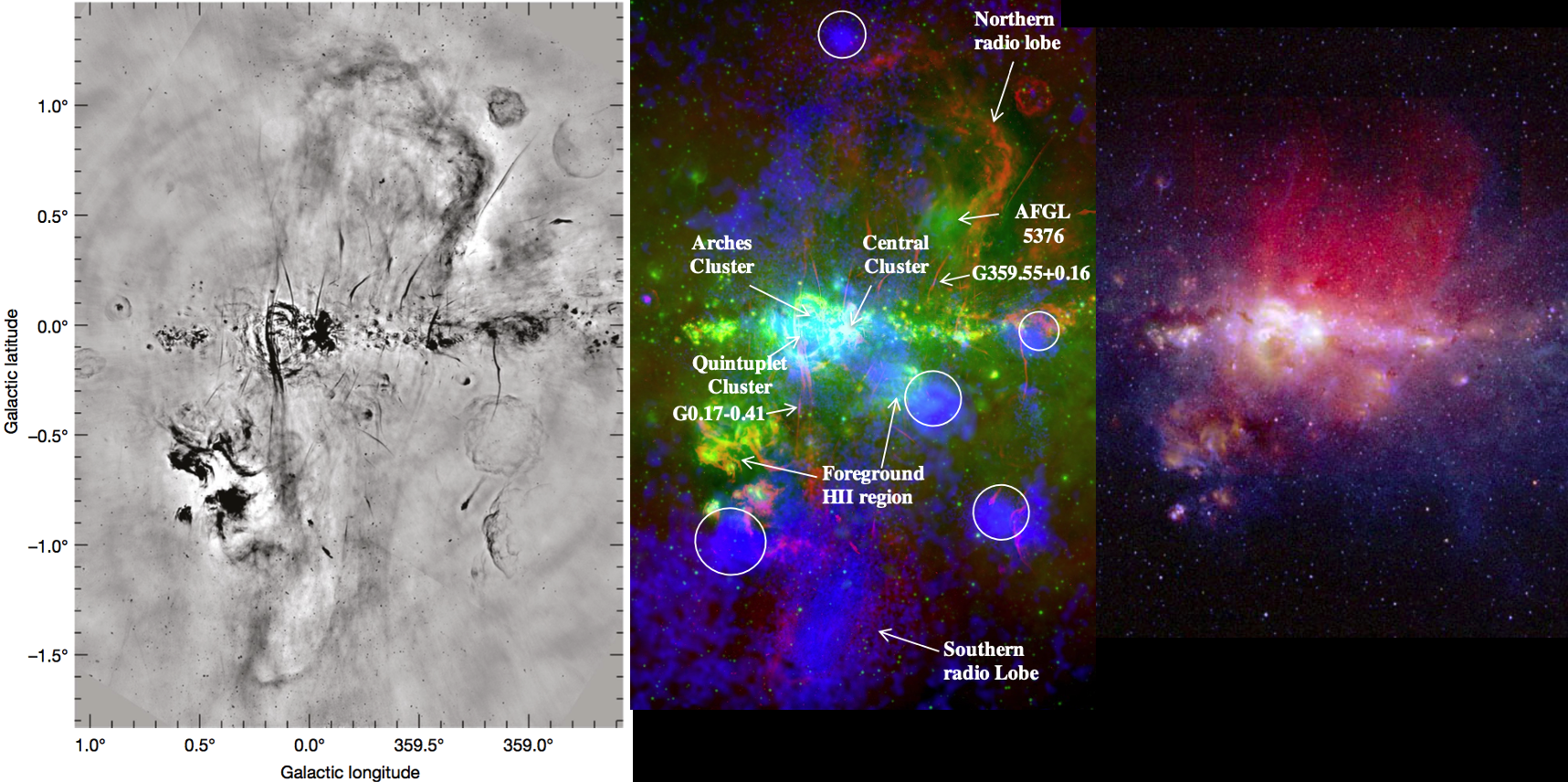}
    \caption{{\bf Left:} 1.28 GHz radio image from MeerKAT \citep{2019Natur.573..235H}.   {\bf Center:}  Radio (red), Mid IR (green), and X-ray (blue) showing the GCL \citep{2021MNRAS.504.1609W}.  White circles mark X-ray artifacts from bright sources.   {\bf Right:}  WISE 22~\um\  (red) and shorter wavelengths (green and blue).  The 3$^o$ long (420 pc) cavity traced by MeerKAT and 22~\um\  dust has an axis of symmetry centered at negative longitudes.  }
    \label{sample-figure}
\end{figure}

The  negative longitude 24~\um\ sources may trace the sheared remnants of a massive cluster which initially contained hundreds of OB stars formed $\sim$10 to 40 Myrs ago (the age of the least massive star to explode as an SN).  Ionization, stellar winds, and supernovae  may have dissociated most molecular clouds at negative longitudes.    How much feedback energy is needed to disrupt a $\sim 10^7$~\Msol\ sector of the CMZ?  The chemical binding energy of \Htwo\ is 4.8 eV.  Thus, the dissociation of $10^7$~\Msol\ of \Htwo\ requires  $\sim >3 \times 10^{52}$~ergs.   The gravitational self-energy of clouds ($\rm  E_G \sim G M^2 / R$)   depends on how the mass is distributed.  If the starburst occurred in a single $10^7$~\Msol\ super-cloud with a radius of $\sim 10 -30$~pc, $E_G \sim 10^{53-54}$ ergs.   If,  the starburst occurred in a hundred smaller  $10^5$~\Msol\ clouds with the same mean density,  $E_G$ would only be $\sim 10^{52-53}$ ergs.   Pre-supernova feedback (ionization, winds, etc.) likely injected as much energy as supernovae.  Thus, to disrupt a sector of the CMZ would require $\sim$100 to 1000 OB stars and supernovae exploding at an average rate of one per $\rm >10^4$ years.     If the initial OB association had an internal velocity dispersion of $\rm \Delta V \sim$10~\kms , over 10 to 40 Myr, shear would have spread the stars over a 100 - 400 pc long arc about the center.  At 100 pc from the center this corresponds to 1/6 to 2/3 of a (circular) orbit.  Today, only a few late O and early B stars may survive and be responsible for the free-free radio emission at negative longitudes. 

Remnants of disrupted molecular gas would today be seen as either low-density thermal HII plasma or neutral HI.  The former would be seen as diffuse free-free radio emission or extended X-ray emission while the latter would be hard to detect given the high opacity of the 21 cm line toward the CMZ.   However, tracers such as CI, or C$^+$ might be used to detect  atomic gas.   In the starburst scenario, feedback  could have blown out an entire sector of the CMZ.   Bubbles created by ionization, stellar winds, and clusters of supernovae  stall as they sweep-up shells in the deep gravitational well of the Galactic center.   The disrupted cloud remnants would remain bound to the CMZ.   However, additional energy supplied by an outburst of the SMBH, accreting stellar remnants such as neutron stars and black holes, or an agent such as shear-amplified magnetic fields and magneto-centrifugal acceleration on a Galactic scale could eventually launch this material into the GCL and the kpc-scale Ferm-LAT bubbles \citep{2018A&A...615A.164V}. 

In summary, I discuss three scenarios to explain the asymmetries:  asymmetric infall, stochastic asymmetry, and a starburst.  Only the latter model can explain both the gas and 24~\um\ asymmetries.  More work is needed to determine the correct explanation for the asymmetries.  We need better measurements of the infall-rates of gas into the CMZ along the leading-edge dust-lanes of the Milky Way's stellar bar.   We need to determine how many gravitationally bound clouds exist in the CMZ.  We need stellar population studies and searches for large amounts of atomic or fully ionized plasma launched out of the Galactic plane from negative longitudes.    Modelers need to incorporate the formation of massive OB associations and stellar-feedback powered super-bubbles in numerical simulations of the CMZ.

\begin{figure}
\center{
    \includegraphics[scale=0.11]{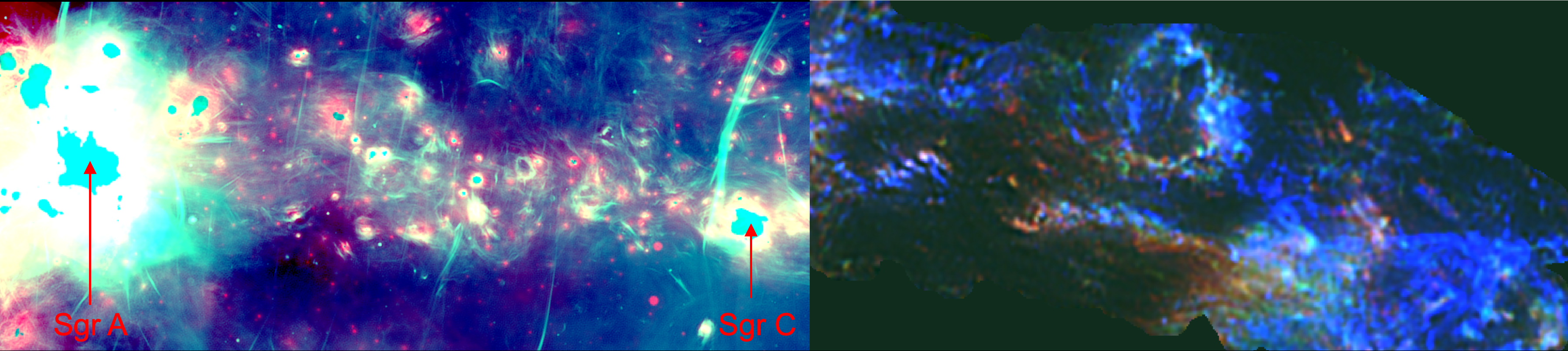}
    \caption{The negative longitude `feedback region' at the base of the GCL which may contain the remnants of a massive OB association formed in a starburst 10 to 40 Myr ago.    {\bf Left:} 1.28 GHz  image from MeerKAT (bluegreen) showing extensive filamentary emission and  the dispersed population of compact 24~\um\ sources (red).   {\bf Right:}  ACES CS 2-1 emission in the same region at -24 to -30 \kms\ (red), -31 to -37 (green), and  -38 to -44 (blue).   Note the ring above the center at [$\sc l,b$]=[359.73$^o$,0.00$^o$].  The panels extend from $\sc l$=359.387$^o$ (right) to 0.016$^o$ (left), $\sc b$=-0.189$^o$ (bottom) to 0.083$^o$ (top).}
    }
    \label{sample-figure}
\end{figure}

\end{document}